\documentclass[sigconf, nonacm]{acmart}

\usepackage{xspace}
\usepackage{pifont}
\usepackage{multirow}

\newcommand{\bfit}[1]{\textbf{\textit{#1}}}
\newcommand{\hi}[1]{\vspace{.25em} \noindent {\bf #1} }
\newcommand{\llm}{\textsc{LLM}\xspace}
\newcommand{\llms}{\textsc{LLMs}\xspace}
\newcommand{\oursys}{\texttt{CrackSQL}\xspace}
\newcommand{\gpt}{\textsf{GPT-4o}\xspace}
\newcommand{\dialectmodel}{\textit{Cross-Dialect Embedding Model}\xspace}

\newcommand{\blue}[1]{\textcolor{blue}{#1}}

\newcommand\vldbdoi{XX.XX/XXX.XX}
\newcommand\vldbpages{XXX-XXX}
\newcommand\vldbvolume{14}
\newcommand\vldbissue{1}
\newcommand\vldbyear{2020}
\newcommand\vldbauthors{\authors}
\newcommand\vldbtitle{\shorttitle} 
\newcommand\vldbavailabilityurl{https://github.com/weAIDB/CrackSQL}
\newcommand\vldbpagestyle{plain} 

\begin{document}
\title{\oursys: A Hybrid SQL Dialect Translation System Powered by Large Language Models}

\author{Wei Zhou}
\affiliation{\institution{{Shanghai Jiao Tong Univ.}}\country{}}
\email{{weizhoudb@gmail.com}}

\author{Yuyang Gao}
\affiliation{\institution{{Tsinghua University}}\country{}}
\email{{21373016@buaa.edu.cn}}

\author{Xuanhe Zhou}
\affiliation{\institution{{Shanghai Jiao Tong Univ.}}\country{}}
\email{{zhouxh@cs.sjtu.edu.cn}}

\author{Guoliang Li}
\affiliation{\institution{{Tsinghua University}}\country{}}
\email{{liguoliang@tsinghua.edu.cn}}

\begin{abstract} 
Dialect translation plays a key role in enabling seamless interaction across heterogeneous database systems. However, translating SQL queries between different dialects (e.g., from PostgreSQL to MySQL) remains a challenging task due to syntactic discrepancies and subtle semantic variations. Existing approaches including manual rewriting, rule-based systems, and large language model (\llm)-based techniques often involve high maintenance effort (e.g., crafting custom translation rules) or produce unreliable results (e.g., \llm generates non-existent functions), especially when handling complex queries.
In this demonstration, we present \oursys, the first hybrid SQL dialect translation system that combines rule and LLM-based methods to overcome these limitations. \oursys leverages the adaptability of LLMs to minimize manual intervention, while enhancing translation accuracy by segmenting lengthy complex SQL via functionality-based query processing. To further improve robustness, it incorporates a novel cross-dialect syntax embedding model for precise syntax alignment, as well as an adaptive local-to-global translation strategy that effectively resolves interdependent query operations. \oursys supports three translation modes and offers multiple deployment and access options including a web console interface, a PyPI package, and a command-line prompt, facilitating adoption across a variety of real-world use cases\footnote{The demonstration video is available at: \blue{\url{https://vimeo.com/1071435762?share=copy}}.}.

\end{abstract}

\maketitle

\pagestyle{\vldbpagestyle}
\begingroup\small\noindent\raggedright\textbf{PVLDB Reference Format:}\\
\vldbauthors. \vldbtitle. PVLDB, \vldbvolume(\vldbissue): \vldbpages, \vldbyear.\\
\href{https://doi.org/\vldbdoi}{doi:\vldbdoi}
\endgroup
\begingroup
\renewcommand\thefootnote{}\footnote{\noindent
This work is licensed under the Creative Commons BY-NC-ND 4.0 International License. Visit \url{https://creativecommons.org/licenses/by-nc-nd/4.0/} to view a copy of this license. For any use beyond those covered by this license, obtain permission by emailing \href{mailto:info@vldb.org}{info@vldb.org}. Copyright is held by the owner/author(s). Publication rights licensed to the VLDB Endowment. \\
\raggedright Proceedings of the VLDB Endowment, Vol. \vldbvolume, No. \vldbissue\ %
ISSN 2150-8097. \\
\href{https://doi.org/\vldbdoi}{doi:\vldbdoi} \\
}\addtocounter{footnote}{-1}\endgroup

\ifdefempty{\vldbavailabilityurl}{}{
\vspace{.3cm}
\begingroup\small\noindent\raggedright\textbf{PVLDB Artifact Availability:}\\
The source code, data, and/or other artifacts have been made available at \blue{\url{\vldbavailabilityurl}}.
\endgroup
}

\section{Introduction}

Dialect translation aims to convert a SQL query written for one database system (e.g., PostgreSQL) into a functionally equivalent query that can be executed on another system (e.g., MySQL). 
This capability is essential for simplifying database migration and supports various applications such as NL2SQL~\cite{NL2SQL}, extract-transform-load (ETL) processes~\cite{babu2016desh}, and cross-database analytics tools~\cite{gavriilidis2023situ}, enabling smooth interaction across different database systems.

\begin{table}[!t]
\vspace{.5cm}
\caption{Comparison of Dialect Translation Methods.}
\label{tab: intro}
\resizebox{\linewidth}{!}{
\begin{tabular}{|l|c|c|c|c|}
\hline
\textbf{Method} & \textbf{\#Dialect} & \textbf{\begin{tabular}[c]{@{}c@{}}Feedback\\ Instruction\end{tabular}} & \textbf{\begin{tabular}[c]{@{}c@{}}Translation\\ Overhead\end{tabular}} & \textbf{\begin{tabular}[c]{@{}c@{}}New Dialect\\ Support\end{tabular}} \\ \hline
\textbf{Human}                                                                     & 1-3 Dialects of Expertise                                       & \checkmark                                                                                              & Minutes                                                                                          & Manual                                                                                        \\ \hline
\textbf{SQLGlot}                                                                     & 24 Common Dialects                                       & $\times$                                                                                              & Seconds                                                                                          & Manual                                                                                        \\ \hline
\textbf{LLM}                                                                         & LLM Compatible                          & $\times$                                                                                              & Seconds to Minutes                                                                                          & Automatic                                                                                     \\ \hline
\textbf{\oursys}                                                                    & 24 + LLM Compatible                          & $\checkmark$                                                                                              & Seconds to Minutes                                                                                          & Automatic                                                                                     \\ \hline
\end{tabular}
}
\end{table}

As summarized in Table~\ref{tab: intro}, existing dialect translation approaches can be classified into three categories:
(1) Manual translation that entirely relies on human expertise across different database dialects for query analysis and rewriting; 
(2) Rule-based methods (e.g., SQLGlot~\cite{sqlglot}) that employ handcrafted translation rules but require extensive custom code for specific cases, resulting in significant maintenance overhead; 
and (3) LLM-based techniques~\cite{DBLP:conf/aidm/NgomK24} that leverage \llm for either direct query translation or rule generation, representing an initial step toward automation. While each approach offers distinct advantages, they all exhibit notable limitations that hinder their effectiveness in practice.

\noindent \textbf{(Limitation 1) High Manual Effort for Code Maintenance.}  
Existing tools (e.g., SQLGlot~\cite{sqlglot}, jOOQ~\cite{jq}, SQLines~\cite{sqlines}) depend on significant manual work by developers to identify equivalent syntax patterns and define corresponding translation rules for different SQL dialects. In addition, maintaining these tools can be challenging, as they often require continuous updates to address translation issues and ensure accuracy. For instance, more than 2,000 unresolved issues remain open in the jOOQ code repository~\cite{jq}, and SQLGlot involves active development with contributions from over 200 developers on GitHub~\cite{sqlglot}.

\noindent \textbf{(Limitation 2) Hallucination Issues in Deterministic Translation.}  
Although \llm demonstrates strong reasoning and coding capabilities that enable it to support new SQL dialects automatically, we observe that hallucination remains a significant issue, particularly with long and complex real-world SQL queries. In such cases, \llm may incorrectly translate queries into operations that are not equivalent or introduce functions that do not exist in the target dialect, often due to limited understanding of dialect-specific features. In addition, it does not make use of valuable external feedback, such as database error messages, to guide or improve the overall translation process.

\begin{figure}[!t]
  \centering
  \includegraphics[width=\linewidth]{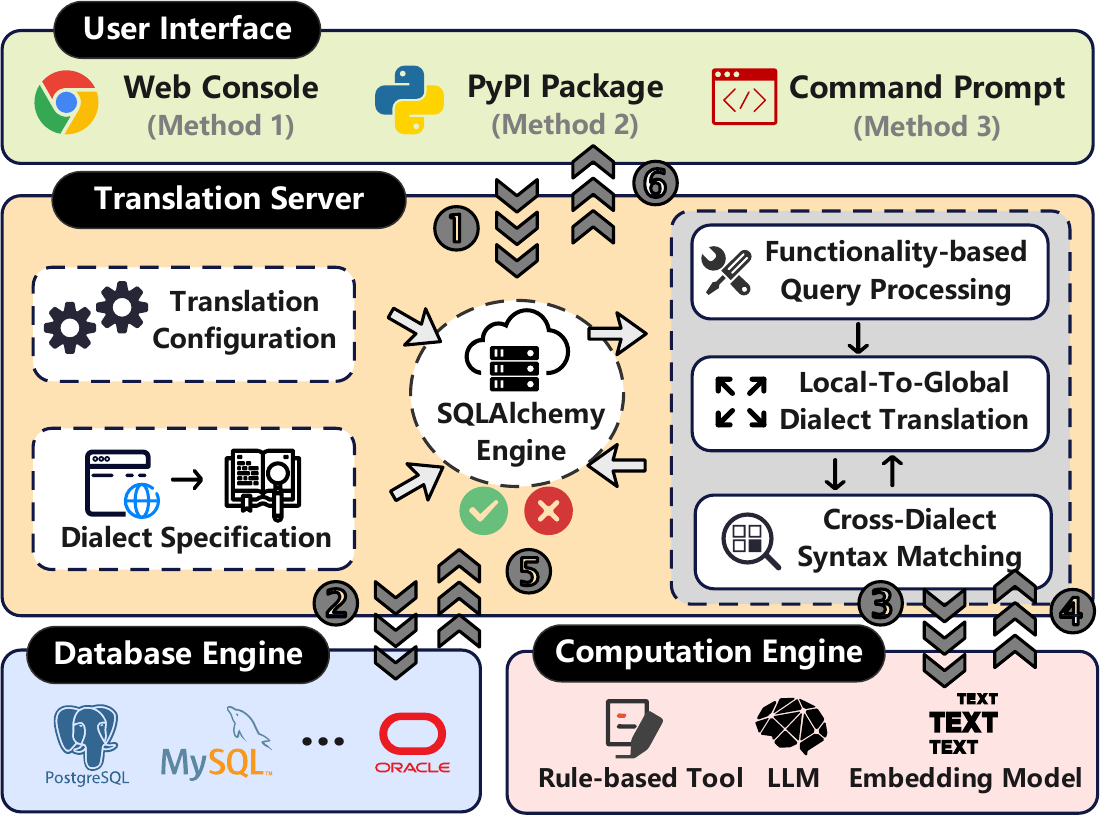}  
  \caption{Architecture of \oursys.}
  \label{fig: cracksql_overview}
\end{figure}

\noindent \textbf{Our Methodology.}  
To address the above limitations, we propose \oursys, the first dialect translation system that integrates both rule-based and \llm-based approaches.
This demonstration is an extension of our research paper~\cite{zhou2025cracksql}. \oursys leverages the ability of \llm to support new SQL dialects automatically, thereby reducing the manual effort required to develop rule-based translation tools. It further improves translation effectiveness by decomposing the input SQL query into smaller functional operations (e.g., the \textsf{CAST()} function), which simplifies the translation process and helps mitigate hallucination issues in long and complex queries.

In addition, \oursys makes the translation process more accurate and consistent through two key techniques:  
(1) A novel cross-dialect embedding model that matches equivalent syntax elements by considering both the code structure (e.g., the BNF tree) of the element and its textual specification, addressing the challenge of identifying equivalent elements across dialects;  
(2) An adaptive translation strategy that incrementally expands the current translation scope, from local-level snippets to broader SQL queries, capturing dependencies among different operations of the query.






\section{\oursys~Overview}
\label{sec: overview}

As illustrated in Figure~\ref{fig: cracksql_overview}, \oursys~comprises four modules to facilitate dialect translation. 
The system supports three distinct operational modes (Table~\ref{tab: mode}), enabling flexible processing of user requests. 
By integrating rule-based methods with \llm, \oursys~ensures robust cross-dialect interoperability while maintaining compatibility with conventional database operations.

\begin{table}[!t]
\caption{Supported Translation Modes in \oursys.}
\label{tab: mode}
\resizebox{\linewidth}{!}{
\begin{tabular}{|l|c|cc|cc|}
\hline
\multicolumn{1}{|c|}{\multirow{2}{*}{\textbf{Mode}}} & \multicolumn{1}{c|}{\multirow{2}{*}{\textbf{\#Dialect}}}      & \multicolumn{2}{c|}{\textbf{\begin{tabular}[c]{@{}c@{}}LLM\\ (w/o \& w fine-tuned)\end{tabular}}} & \multicolumn{2}{c|}{\textbf{\begin{tabular}[c]{@{}c@{}}Embedding Model\\ (w/o \& w fine-tuned)\end{tabular}}} \\ \cline{3-6} 
\multicolumn{1}{|c|}{}                               & \multicolumn{1}{c|}{}                                         & \multicolumn{1}{c|}{\textbf{Cloud}}                   & \multicolumn{1}{c|}{\textbf{Local}}       & \multicolumn{1}{c|}{\textbf{Cloud}}                         & \multicolumn{1}{c|}{\textbf{Local}}             \\ \hline
\textbf{Rule-only}                                                                                   & 24                                                                                                                              & \multicolumn{1}{c|}{$\times$}                                                                        & $\times$                                                & \multicolumn{1}{c|}{$\times$}                                                                              & $\times$                                                      \\ \hline
\textbf{LLM-direct}                                                                                  & \checkmark                                                                                                       & \multicolumn{1}{c|}\checkmark                                                & \checkmark                        & \multicolumn{1}{c|}{$\times$}                                                                              & $\times$                                                      \\ \hline
\textbf{Rule+LLM}                                                                                    & \begin{tabular}[c]{@{}c@{}}3\\ (PG/MySQL/Oracle)\end{tabular}                                                                   & \multicolumn{1}{c|}\checkmark                                                & \checkmark                        & \multicolumn{1}{c|}\checkmark                                                      & \checkmark                              \\ \hline
\end{tabular}
}
\end{table}

\noindent \textbf{Workflow.}
The translation process begins when the translation requests are submitted via the \emph{User Interface} with three methods (i.e., the web console, the PyPI package\footnote{The PyPI package can be found at: \blue{\url{https://pypi.org/project/cracksql/}}.}, and the command prompt).
Then, these requests are routed to the unified \emph{Translation Server}, which orchestrates the core translation logic via the SQLAlchemy engine (\ding{182}).
The server dynamically configures translations based on user-specified parameters, with different operational modes requiring distinct services.
For example, the \textsf{Rule + \llm} mode conducts translation in the following steps (the detailed technical design are presented in Section~\ref{sec: method}).
It first undergoes functionality-based query processing to decompose the whole SQL into several snippets for ease of \llm translation over lengthy SQL.
Then, it employs local-to-global translation strategy instructed by the \emph{Database Engine}, which locates the prepared SQL snippets with execution errors (\ding{183}).
During the translation process, it will iteratively turn to \emph{Computation Engine} in multiple rounds with two basic operations (\ding{184}):
(1) undergo hybrid translation combining rule-based tools or \llm over the erroneous SQL snippets;
(2) perform cross-dialect syntax matching based on our newly-designed embedding model to enrich the context for \llm with relevant dialect specifications (e.g., the usage descriptions of two equivalent functions).
The translated SQL snippets are then returned to the server (\ding{185}) and passed to the \emph{Database Engine} to check for any remaining execution errors (\ding{186}).
If no errors are found, the successfully translated SQL snippets are then forwarded to the \emph{User Interface} (\ding{187}).

\section{Methodology}
\label{sec: method}

\oursys~proposes the following techniques with rule and \llm-based methods to address the challenges in \llm-based translation.

\hi{Functionality-based Query Processing.}
\llms frequently encounter hallucination issues when processing lengthy queries characterized by extensive content and complex structural syntax elements, particularly those involving the translation of multiple syntactic specifications.
Thus, we propose the following strategies.

\noindent $\bullet$ \bfit{Tree-based Segmentation.}
Given the syntax tree, the original query \( Q^O \) is decomposed into distinct functional operations in the source dialect (e.g., MySQL \textsf{TIMESTAMPDIFF()} function). 
Thus, the complex query is segmented into a set of simpler operations \( \{q'_i\} \). 

\noindent $\bullet$ \bfit{Rule-based Simplification \& Abstraction.}
Each segmented operation is further simplified through specific rule-based approaches, such as operation normalization (converting custom functions into standard equivalents) and operation abstraction (substituting translation-irrelevant details like column or table names with generic non-terminal symbols). 

This entire query processing workflow is carried out in an automatic manner, involving parsing and traversing the SQLs, and aligning them with predefined specifications through a BNF grammar parser using a depth-first traversal strategy.

\hi{Cross-Dialect Syntax Matching.}
To identify syntactically equivalent elements across different SQL dialects, we propose a \dialectmodel that generates functional embeddings to support accurate syntax matching. This model is trained using a novel \emph{Retrieval-Enhanced Contrastive Learning}, which utilizes carefully constructed positive and negative pairs of syntax elements with equivalent and distinct functionalities, respectively.

\begin{figure*}[!t]
\vspace{-1.cm}
  \centering
  \includegraphics[width=\linewidth]{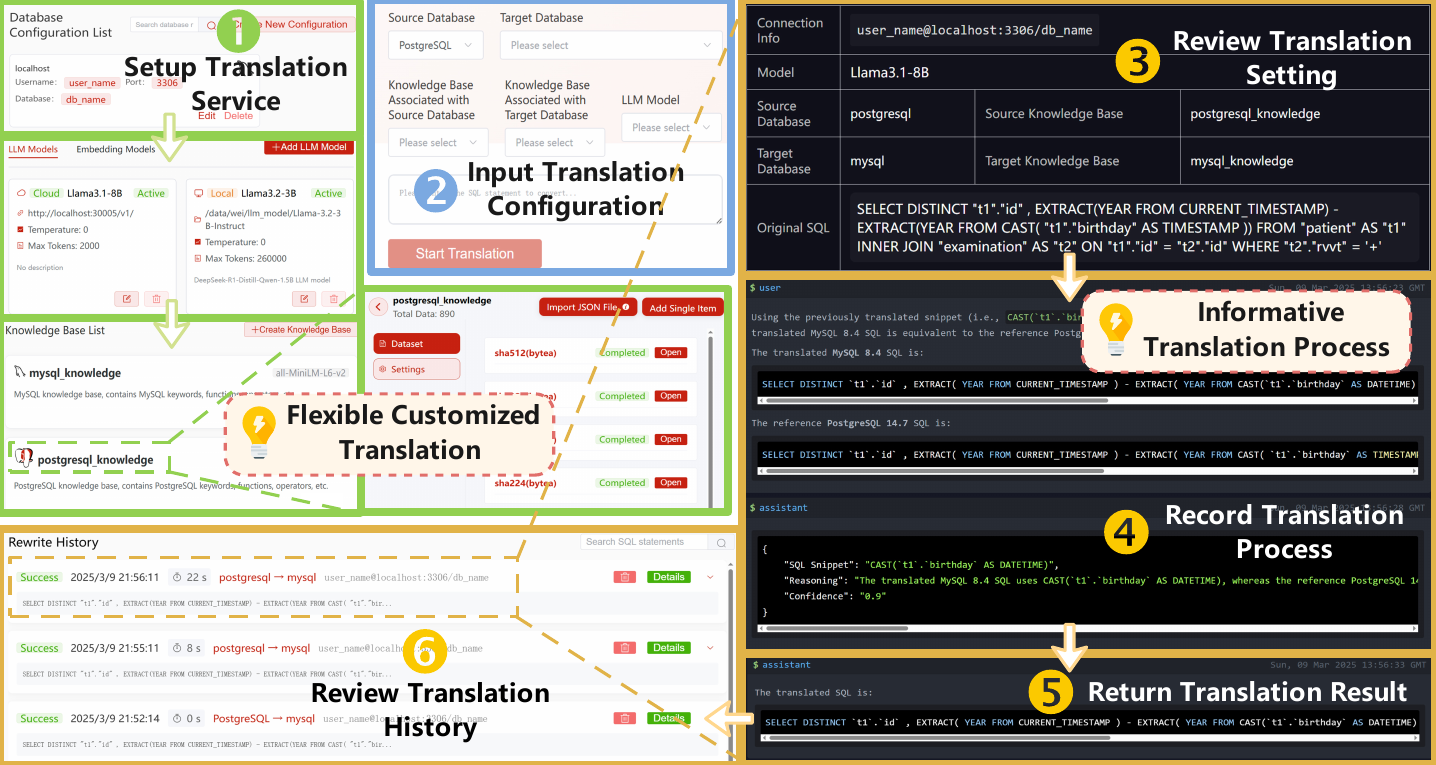}  
  \caption{Demonstration of \oursys (Users submit translation requests through web console interface).}
  \label{fig: demo}
\end{figure*}

\noindent $\bullet$ \bfit{\dialectmodel.} The proposed embedding model consists of three key components:
(1) Code structure encoding, which captures the hierarchical and structural information of syntax elements based on their representation in abstract syntax trees;
(2) Syntax specification encoding using a Mixture-of-Experts (MoE) approach, allowing the model to flexibly handle syntax definitions of varying lengths and complexities; and
(3) Syntax-specification aggregation, which aligns the encoded textual syntax specification descriptions with corresponding syntax elements of code structure to produce unified embeddings for matching.

\noindent $\bullet$ \bfit{Retrieval-Enhanced Contrastive Learning.}
To train the embedding model effectively, we design diverse strategies to construct meaningful positive and negative training examples.
Positive examples include (1) augmented versions of a syntax element’s own specification (e.g., via synonym substitution), (2) syntax elements with the same keywords and functionalities across different dialects, and (3) equivalent pairs identified by existing rule-based translation tools. 
For negative examples, we refine random sampling methods to produce hard negatives, i.e., pairs of syntax elements that are functionally different but semantically similar. These hard negatives are included alongside positive examples within the same training batch, encouraging the model to accurately differentiate truly equivalent syntax elements from those with superficial similarities.

\hi{Local-To-Global Dialect Translation.} 
To prevent unnecessary translations and facilitate simultaneous consideration of multiple syntax elements in complex translations, we propose a hierarchical translation approach. 
Specifically, it translates the processed operations \(\{q'_i\}\) into the final query \(Q^T\) in the target dialect by progressively extending query operations from local to global contexts.

\noindent $\bullet$ \bfit{Focused Local-Level Translation.}
At the local operation level, we identify problematic operations (\(\{q'_j\}\)) that lead to translation failures, which are categorized as: (1) incompatibilities, detected by the BNF parser due to syntax mismatches, and (2) insufficiencies, occurring when translation attempts fail due to incomplete local syntax information (e.g., exceeding the maximum allowed trials of the \llm).
For operations identified as incompatible, we employ the \dialectmodel to match equivalent syntax elements in the target dialect, and subsequently translate these operations using a hybrid strategy that integrates rule-based translation tools (e.g., SQLGlot) and \llm-based translation.

\noindent $\bullet$ \bfit{Adaptive Global-Level Expansion.}
For insufficiently specified operations, we iteratively extend the operation scope by incorporating adjacent syntax elements, resulting in extended operations (\(q''_j\)). This iterative translation process continues until the translated operations collectively replace the original SQL snippets, producing a final query \(Q^T\) that is both syntactically valid and functionally equivalent to the original query \(Q^O\).

Note that the usage of \llm translation is limited to specific segmented SQL operations corresponding to predefined functionality within prepared specifications from official documents, and \llm hallucinations are mitigated through hybrid validation mechanisms combining rule-based parser checking and \llm-based reflection.

\section{Demonstration}
\label{sec: demo}

We demonstrate the translation accuracy of \oursys on real-world SQL queries and showcase how users can perform translations through an interactive web console designed for ease of use. In addition, we provide a Python package for local or programmatic access, which is available at: \blue{\url{https://pypi.org/project/cracksql/}}.

\subsection{Web Console Interface}
\label{subsec: demo}

The translation process begins with an initial setup, during which users specify a set of parameters to configure the translation service. In particular, users define: (1) the target databases used to provide feedback during translation, (2) the model components, including the embedding model for syntax matching and the \llm for translating SQL snippets, and (3) the external knowledge base used to augment the \llm's performance (\ding{182}).
After the setup, users input the SQL query to be translated, choose the desired translation configurations, and initiate the process by clicking the ``Start Translation'' button (\ding{183}). This triggers the creation of a background translation task, through which users can: (1) review the configured translation settings (\ding{184}), (2) inspect the recorded translation process for individual SQL snippets (\ding{185}), and (3) access the final translated SQL query along with the result (\ding{186}).
Furthermore, the system maintains a history of translation requests, allowing users to browse past translations in the translation detail list (\ding{187}).

\noindent \textbf{Summary.}  
\oursys offers \emph{Flexible Customized Translation}, enabling users to easily adjust translation settings through a user-friendly interface. In addition, \oursys provides an \emph{Informative Translation Process} that enables users to gain deeper insights into the translation workflow and identify potential issues that may arise during execution.

\subsection{Experiment Result}
\label{subsec: exp}

\begin{figure}[!t]
  \centering
  \includegraphics[width=\linewidth]{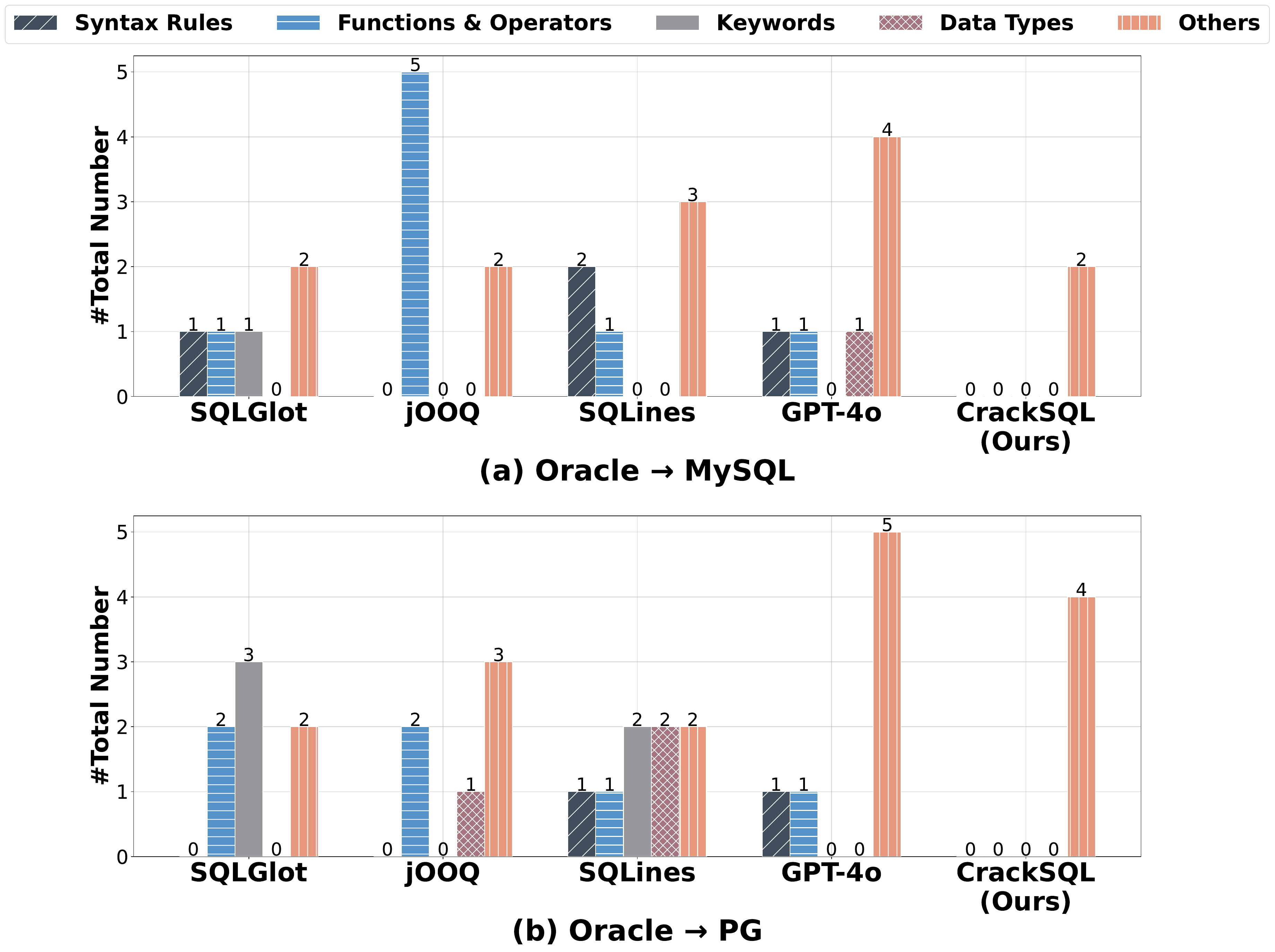}  
  \caption{Error Distribution of Different Methods.}
  \label{fig: error_num}
\end{figure}

Figure~\ref{fig: error_num} presents the evaluation results on a modified version of the real-world BIRD benchmark, focusing on dialect translations from $Oracle \rightarrow MySQL$ and $Oracle \rightarrow PostgreSQL$.

\begin{sloppypar}
\noindent \textbf{(Observation 1) \oursys~consistently outperforms other methods with fewer errors introduced in the translated SQLs.} 
 Figure~\ref{fig: error_num} shows that \oursys~introduces zero errors in four out of the five error categories. Specifically, we find no instances of syntax-rule violations, incorrect function or operator mappings, keyword misuse, or mismatched data types when using \oursys. This can be attributed to its two primary designs: (1)~\llm-based code generation that actively incorporates dialect specifications retrieved through the cross-dialect embedding model (enabling the system to auto-detect and adapt to the correct syntax for each dialect), and (2)~the functionality-based query processing alongside a local-to-global translation strategy. By segmenting queries into smaller functional components and only resorting to the LLM for targeted sub-operations, \oursys~reduces the chance of \llm hallucination and ensures robust handling of nested or complex SQL queries. These combined features substantially improve translation coverage and correctness relative to the other baselines, which rely heavily on manually curated rules or naive LLM prompting without dialect-level constraints.
\end{sloppypar}

\noindent \textbf{(Observation 2) Rule-based methods exhibit inconsistent error distributions across different translation scenarios.} Although rule-based translators (e.g., SQLGlot, jOOQ, and SQLines) sometimes perform comparably to \oursys~for specific operations, their overall error patterns vary significantly across tasks like $Oracle \rightarrow MySQL$ versus $Oracle \rightarrow PostgreSQL$. For instance, jOOQ introduces more ``Functions \& Operators'' errors when translating certain date/time calculations from Oracle to MySQL, whereas SQLGlot misuses or omits some keywords during $Oracle \rightarrow PostgreSQL$ conversions. These inconsistencies stem from each tool’s coverage gaps in hand-crafted mappings: if the relevant function or keyword equivalences have not been explicitly encoded, the translation either fails outright or produces incorrect results. In contrast, \oursys~systematically references target-dialect syntax specifications through the cross-dialect embedding model, resulting in more uniform accuracy gains even when the operation is uncommon or spans multiple SQL clauses.

\begin{sloppypar}
\noindent \textbf{(Observation 3) Direct usage of large language models can lead to additional translation errors due to hallucination.} When we directly prompt \gpt to translate Oracle queries to MySQL or PostgreSQL, we observe an increase in ``Other Errors’’—namely, incorrect column renaming, unintended data type changes (e.g., using types from a different dialect), or spurious clauses that do not appear in the source query. Such mistakes tend to arise from general-purpose LLM behavior that is effective for open-ended text generation but insufficiently constrained for SQL translation, which demands exact syntactic and semantic fidelity. Consequently, while \gpt can produce successful translations in many straightforward cases, it frequently fails on edge cases that require close adherence to version-specific SQL standards or specialized operators. By contrast, \oursys~limits LLM-generated transformations to precisely those query parts that truly need rewriting, validates each translated snippet using the target dialect’s parser, and cross-checks any uncertain operation via the embedded specifications. This meticulous workflow greatly reduces the risk of hallucination errors and boosts the final translation accuracy.
\end{sloppypar}


\vspace{.25cm}

\bibliographystyle{ACM-Reference-Format}
\bibliography{sample}


\begin{thebibliography}{8}


\ifx \showCODEN    \undefined \def \showCODEN     #1{\unskip}     \fi
\ifx \showDOI      \undefined \def \showDOI       #1{#1}\fi
\ifx \showISBNx    \undefined \def \showISBNx     #1{\unskip}     \fi
\ifx \showISBNxiii \undefined \def \showISBNxiii  #1{\unskip}     \fi
\ifx \showISSN     \undefined \def \showISSN      #1{\unskip}     \fi
\ifx \showLCCN     \undefined \def \showLCCN      #1{\unskip}     \fi
\ifx \shownote     \undefined \def \shownote      #1{#1}          \fi
\ifx \showarticletitle \undefined \def \showarticletitle #1{#1}   \fi
\ifx \showURL      \undefined \def \showURL       {\relax}        \fi
\providecommand\bibfield[2]{#2}
\providecommand\bibinfo[2]{#2}
\providecommand\natexlab[1]{#1}
\providecommand\showeprint[2][]{arXiv:#2}

\bibitem[\protect\citeauthoryear{??}{sql}{[n.d.]a}]%
        {sqlglot}
 \bibinfo{year}{[n.d.]}\natexlab{a}.
\newblock \showarticletitle{https://sqlglot.com/sqlglot.html}.
\newblock
\newblock
\shownote{Last accessed on 2024-10.}


\bibitem[\protect\citeauthoryear{??}{jq}{[n.d.]}]%
        {jq}
 \bibinfo{year}{[n.d.]}\natexlab{}.
\newblock \showarticletitle{https://www.jooq.org/}.
\newblock
\newblock
\shownote{Last accessed on 2024-10.}


\bibitem[\protect\citeauthoryear{??}{sql}{[n.d.]b}]%
        {sqlines}
 \bibinfo{year}{[n.d.]}\natexlab{b}.
\newblock \showarticletitle{https://www.sqlines.com/}.
\newblock
\newblock
\shownote{Last accessed on 2024-10.}


\bibitem[\protect\citeauthoryear{Babu and Gunasingh}{Babu and Gunasingh}{2016}]%
        {babu2016desh}
\bibfield{author}{\bibinfo{person}{Chitra Babu} {and} \bibinfo{person}{G Gunasingh}.} \bibinfo{year}{2016}\natexlab{}.
\newblock \showarticletitle{DESH: Database evaluation system with hibernate ORM framework}. In \bibinfo{booktitle}{\emph{2016 International Conference on Advances in Computing, Communications and Informatics (ICACCI)}}. IEEE, \bibinfo{pages}{2549--2556}.
\newblock


\bibitem[\protect\citeauthoryear{Gavriilidis, Beedkar, Quian{\'e}-Ruiz, and Markl}{Gavriilidis et~al\mbox{.}}{2023}]%
        {gavriilidis2023situ}
\bibfield{author}{\bibinfo{person}{Haralampos Gavriilidis}, \bibinfo{person}{Kaustubh Beedkar}, \bibinfo{person}{Jorge-Arnulfo Quian{\'e}-Ruiz}, {and} \bibinfo{person}{Volker Markl}.} \bibinfo{year}{2023}\natexlab{}.
\newblock \showarticletitle{In-situ cross-database query processing}. In \bibinfo{booktitle}{\emph{2023 IEEE 39th International Conference on Data Engineering (ICDE)}}. IEEE, \bibinfo{pages}{2794--2807}.
\newblock


\bibitem[\protect\citeauthoryear{Kim, So, Han, and Lee}{Kim et~al\mbox{.}}{2020}]%
        {NL2SQL}
\bibfield{author}{\bibinfo{person}{Hyeonji Kim}, \bibinfo{person}{Byeong{-}Hoon So}, \bibinfo{person}{Wook{-}Shin Han}, {and} \bibinfo{person}{Hongrae Lee}.} \bibinfo{year}{2020}\natexlab{}.
\newblock \showarticletitle{Natural language to {SQL:} Where are we today?}
\newblock \bibinfo{journal}{\emph{Proc. {VLDB} Endow.}} \bibinfo{volume}{13}, \bibinfo{number}{10} (\bibinfo{year}{2020}), \bibinfo{pages}{1737--1750}.
\newblock


\bibitem[\protect\citeauthoryear{Ngom and Kraska}{Ngom and Kraska}{2024}]%
        {DBLP:conf/aidm/NgomK24}
\bibfield{author}{\bibinfo{person}{Amadou~Latyr Ngom} {and} \bibinfo{person}{Tim Kraska}.} \bibinfo{year}{2024}\natexlab{}.
\newblock \showarticletitle{Mallet: {SQL} Dialect Translation with {LLM} Rule Generation}. In \bibinfo{booktitle}{\emph{aiDM 2024}}. \bibinfo{publisher}{{ACM}}, \bibinfo{pages}{3:1--3:5}.
\newblock
\urldef\tempurl%
\url{https://doi.org/10.1145/3663742.3663973}
\showDOI{\tempurl}


\bibitem[\protect\citeauthoryear{Zhou, Gao, Zhou, and Li}{Zhou et~al\mbox{.}}{2025}]%
        {zhou2025cracksql}
\bibfield{author}{\bibinfo{person}{Wei Zhou}, \bibinfo{person}{Yuyang Gao}, \bibinfo{person}{Xuanhe Zhou}, {and} \bibinfo{person}{Guoliang Li}.} \bibinfo{year}{2025}\natexlab{}.
\newblock \showarticletitle{{Cracking SQL Barriers:} {An} LLM-based Dialect Transaltion System}.
\newblock \bibinfo{journal}{\emph{Proc. {ACM} Manag. Data}} \bibinfo{volume}{3}, \bibinfo{number}{3} (\bibinfo{year}{2025}).
\newblock


\end{thebibliography}

\end{document}